\documentclass[twocolumn,nofootinbib,superscriptaddress]{revtex4-1}
\usepackage[utf8]{inputenc}
\usepackage{longtable}
\usepackage{graphicx}
\usepackage{hyperref} 
\usepackage{placeins}
\usepackage{mathrsfs,amsmath,amssymb}   
\usepackage{dsfont} 

\newcommand{\eq}[1]{\begin{equation}#1\end{equation}}
\newcommand{\al}[1]{\begin{align}#1\end{align}}
\newcommand{\Q}{\mathcal{Q}}
\newcommand{\R}{\mathcal{R}}
\renewcommand{\S}{\mathcal{S}} 
\newcommand{\C}{\mathcal{C}}
\newcommand{\E}{\mathcal{E}}
\newcommand{\F}{\mathscr{F}}

\newcommand{\mC}{\mathbf{C}}
\newcommand{\mS}{\mathbf{S}}
\newcommand{\mk}{\mathbf{k}}
\newcommand{\mg}{\mathbf{g}}

\newcommand{\LR}[2]{\left #1 \rule{0cm}{#2}\right.}
\newcommand{\BBL}{\LR{(}{.7cm}}
\newcommand{\BBR}{\LR{)}{.7cm}}

\begin{document}


\title{Nonlinear price impact from linear models}

\author{Felix Patzelt}
\email{felix@neuro.uni-bremen.de}
\affiliation{Capital Fund Management, 23 rue de l'Universit{\'e}, 75007, Paris, France}
\affiliation{Supported by Deutsche Forschungsgemeinschaft}
\author{Jean-Philippe Bouchaud}
\affiliation{Capital Fund Management, 23 rue de l'Universit{\'e}, 75007, Paris, France}

\begin{abstract}
The impact of trades on asset prices is a crucial aspect of market dynamics for academics, regulators and practitioners alike. Recently, universal and highly nonlinear master curves were observed for price impacts aggregated on all intra-day scales \cite{patzelt2017universal}. Here we investigate how well these curves, their scaling, and the underlying return dynamics are captured by linear ``propagator'' models. We find that the classification of trades as price-changing versus non-price-changing can explain the price impact nonlinearities and short-term return dynamics to a very high degree. The explanatory power provided by the change indicator in addition to the order sign history increases with increasing tick size. 
To obtain these results, several long-standing technical issues for model calibration and -testing are addressed. We present new spectral estimators for two- and three-point cross-correlations, removing the need for previously used approximations. We also show when calibration is unbiased and how to accurately reveal previously overlooked biases. Therefore, our results contribute significantly to understanding both recent empirical results and the properties of a popular class of impact models.
\end{abstract}

\maketitle

\tableofcontents


\section{Introduction}

According to economic theory, prices should reflect information. In practice, price formation takes place during trading, which presumably contributes to incorporating said information into the price \cite{kylei1985continuous, fama1970efficientMarkets, lyons2000microstructure}. Empirically, buying or selling an asset tends to push prices up or down \cite{hopman2007prices, Bouchaud2009MarketsSlowlyDigest}. For active market participants like large institutions, this price impact contributes significantly to total transaction costs. Despite its importance, however, price impact is not yet fully understood.

In modern electronic markets, prices are formed in a continuous double auction via a Limit-Order Book (LOB). Traders can either submit \emph{limit orders} to buy or sell a certain volume at a fixed price or \emph{market orders} that get executed immediately at the best available price. Limit orders \emph{provide liquidity} by filling the book with offers that are executed at a later point in time. Market orders \emph{take liquidity} by triggering transactions that remove existing offers from the book. 

Because the immediately available liquidity at a certain price is limited, trades with large volumes are typically fragmented. Such a sequence of incrementally executed transactions following a single decision to trade a large volume is called a \emph{metaorder}. It has been suggested that the execution of metaorders is responsible for a peculiar observation: the \emph{signs} of market orders (buy: $\epsilon = +1$, sell: $\epsilon = -1$) are positively correlated over long periods of time \cite{Bouchaud2009MarketsSlowlyDigest}. Here, we investigate four models, three of which were originally introduced to reconcile these long-ranged order sign correlations with price returns being almost diffusive (first-order uncorrelated). Our main objective, however, is to test their consistency with some very recent empirical findings.

In \cite{patzelt2017universal}, we revisited the question how $N$ subsequent trades impact the price. We found a universal sigmoidal shape for the expected return conditioned on the \emph{volume imbalance}. That is, the difference between the volumes of all buy- and all sell-market-orders. These master curves were recovered by finding the proper rescaling with $N$, close to the Hurst exponents for the returns and order-signs.
Moreover, the \emph{aggregate sign impact} was found to have an sinusoidal shape: when the order flow is strongly polarised in one direction, the expected return is close to zero \cite{patzelt2017universal}. This effect was traced back to prices being pinned to a level where liquidity provision completely offsets the extreme bias of the arriving market orders. In fact, the probability that a market order changes the mid-price was found to be a decreasing function of the magnitude of the order-sign bias. These results correct previous reports \cite{Bouchaud2009MarketsSlowlyDigest} that the highly concave single-trade impact became more linear on aggregation. Hence aggregate impact joins the well-known square-root impact of metaorders (see e.g. \cite{toth2011anomalous} and the references therein) as a challenge to linear impact models like the classical Kyle model \cite{kylei1985continuous}.

In the present paper, we investigate the capability of four closely related models to reproduce these effects. They received recent attention in the literature and belong to a model class that is popular among practitioners. We consider the real (market-) order flow as an external input and model its effect on future returns. More precisely, we consider each trade $t$ an event and characterise it by the sign $\epsilon(t)$ of the corresponding market order and a label $\pi(t)$ indicating its event-type: price-changing ($\pi(t)=c$) or non-price-changing ($\pi(t)=n$). The generation of synthetic order flows is left for future work.

As stated above, the original aim of three of the considered models was to remove the strong correlations injected by the order flow from the models' output returns (see e.g. \cite{eisler2011models, eisler2012price, taranto2016linear}). The underlying insight is that this would be impossible if the impact of a trade were permanent \emph{and} time-invariant. This led to the classification as either Transient Impact Models (TIMs) or History Dependent Impact Models (HDIMs). As we will see in their full definition in section \ref{sec:models}, however, all models express the return at time $t$ as a linear combination of past order-signs $\epsilon(t')$ by means of a convolution with one or several ``propagator'' kernels. The formal differences, which lie exclusively in the way the labels $\pi(t')$ influence each models output, are actually more important for the purpose of this paper than their conventional interpretations. We furthermore introduce a Constant Impact Model (CIM), which only depends on the current state, as a limiting case. It can also be seen as a reduction of the Constant Gap Model \cite{eisler2011models} to two event types in trade time.

Our main result is that the labels $\pi$ are most important to model price impact. In fact, the HDIM2 reproduces \emph{all} observations very well.\footnote{The suffix ``$2$'' indicates that this model uses two input signals} This includes the detailed shape of the sigmoidal and sinusoidal impact functions and the signature plot, which quantifies price diffusion for different lags. This agreement is made possible thanks to a new, exact calibration procedure that requires the estimation of 3-point correlation functions. In contrast, previous work on HDIMs used an approximation involving 2-point correlations, which may lead to spurious conclusions. 

The CIM2 performs almost as well as the HDIM2 for some observables, but exhibits long-term superdiffusive prices especially for small-tick instruments. The TIMs, on the other hand, fare quite well for small-tick sizes, but exhibit severe problems when the tick size is increased. Analysing how the models' performances depend on the available information allows us to develop a detailed understanding of the factors determining aggregate price impact.

\section{The Data}
\label{sec:data}

We used the same data set as in \cite{patzelt2017universal}:
\begin{itemize}
\item 12 technology stocks on the US primary NASDAQ market, for the years 2011 to 2016. This includes some of the most traded stocks in the world like Apple (AAPL) and Microsoft (MSFT).
\item the 13 highest turnover stocks on NASDAQ OMX NORDIC (called just OMX in the following), which covers the Nordic markets Stockholm, Helsinki, and Copenhagen for October 2011 until end of September 2015. OMX is the primary market for the selected stocks.
\item 6 futures on EUREX EBS (BOBL, BUND, DAX, EUROSTOXX, SCHATZ, SMI) for October 2014 until the end of 2015.
\end{itemize}
and the same code for preprocessing and cleaning. The main features are summarised below; see \cite{patzelt2017universal} for further discussion.

The sample offers a wide range of structural diversity. We found that particularly the effects of price-discretisation play an important role which can be quantified by the microstructural parameter \cite{dayri2015large, huang2015predict}:
\eq{
	\label{eq:eta}
	\eta := \frac{N_c}{2 N_a}
}
where $N_c$ is the number of subsequent price movements in same direction (continuations) and $N_a$ the number of price-movements in alternating directions. It measures the effect of discretisation of a diffusion process. $\eta > 0.5$ corresponds to small-tick instruments and $\eta < 0.5$ to large-tick instruments.

Prices on NASDAQ are discretised with a fixed tick size of $\$0.01$, which can be considered very small ($\eta = 0.73$) to medium ($\eta = 0.49$) for the analysed stocks. Up to roughly one third of the transactions were executed against hidden liquidity. Orders are automatically routed to a different market within the US when a better offer is available.
On OMX, market fragmentation is typically lower. Tick sizes vary with price and are effectively larger ($0.24 \leq \eta \leq 0.50$) than for NASDAQ. Here, hidden liquidity represents a vanishingly small fraction of all traded volume. Finally, the EUREX futures are not traded on other platforms. Tick sizes vary between moderately large ($\eta = 0.44$) and extremely large ($\eta =0.03$).

We calculate price-returns $r(t) = \log m({t+1}) - \log m(t)$ from the mid-prices $m$ defined as the average of the bid price and the ask price just before each trade. Order-signs were reconstructed by by labelling all trades above the mid-price as $\epsilon=+1$ and all trades below as $\epsilon=-1$. Trades exactly at the mid-price were discarded, as were obviously irregular entries such as transactions labelled as irregular by the exchange or provider or entries with non-finite prices (including bid and ask). Transactions with the same sign and millisecond timestamp were merged. Finally, we constructed the event-type labels
\eq{
	\label{eq:change_indicator}
	\pi(t) = \left \{ \begin{array}{lcl}
		n \quad & \mathrm{if} & \quad  m(t+1) = m(t)\\
		c & \mathrm{else} &
	\end{array} \right.
}
to indicate whether a trade changed the mid-price.

For each trading day, we analysed only trades between 30 minutes after opening and before closing. Days with shortened trading-hours were discarded. Since we limited our study to intra-day data, ignoring overnight price changes, we calculated all statistics on a day-by-day basis and then averaged. Maximum lengths for bins, correlation- and kernel lags were chosen as the length of the shortest day for the respective instrument and time period. We used overlapping bins for the daily estimates on longer days to avoid unnecessarily discarding data. We also simulated the models for each day separately, such that no information from the previous day ``spilled over''. Nevertheless, we also tested--where possible--calibration and simulation while treating all days as a single time series. We found only minor quantitative differences.

\section{The Models}
\label{sec:models}

Propagator models express the mid-price return $r(t)$ at time $t$ as a linear combination of the previous order-signs $\epsilon(t')$ at $t' \leq t$, weighted by kernels that possibly depend on the corresponding event type $\pi$. 

The simplest case with only one kernel is the TIM1 \cite{bouchaud2004fluctuations} with
\eq{
	\label{eq:TIM1}
	r_{\mathrm{TIM1}}(t) := \sum\limits_{j \geq 0} g(j)\, \epsilon(t-j)
}
where we call $g$ the (differential) kernel or ``propagator''.

The TIM2 expresses returns as the sum of \emph{two} convolutions: one with kernel $g_c$ for signs of price-changing events and another one with kernel $g_n$ for non-price-changing events:
\eq{
	\label{eq:TIM2}
	r_{\mathrm{TIM2}}(t) := \sum\limits_{\pi'}  \sum\limits_{j \geq 0} g_{\pi'}(j)\ \delta_{\pi(t-j)\, \pi'}\ \epsilon(t-j)\\
}
where $\delta$ is the Kronecker delta and the first sum runs over all considered event types $\pi' \in \{n, c\}$. Note that non-zero $r_{\mathrm{TIM2}}(t)$ are likely even when $\pi(t) = n$, since the decay of past events is always affecting the price in this model.

The HDIM2 mends this inconsistency by switching between four kernels $\kappa_{\pi' \pi''}$ based also on the label of the  latest event:
\eq{
	\label{eq:HDIM2}
	r_{\mathrm{HDIM2}}(t) := \sum\limits_{\pi''} \delta_{\pi(t)\, \pi''} \sum\limits_{\pi'}  \sum\limits_{j \geq 0} \kappa_{\pi' \pi''}(j)\ \delta_{\pi(t-j)\, \pi'}\ \epsilon(t-j)\\
}
where $\kappa_{\pi' \pi''}(j) := \kappa_{\pi'}(j)\, \delta_{\pi'' c}$ ensures consistency with Eq.~\ref{eq:change_indicator}. In other words, zero returns for accordingly labelled events are enforced by definition, effectively allowing for only two kernels with finite coefficients. Note that $\kappa_{\pi' \pi''}(0)$ is only ever is used for $\pi' = \pi''$.

Finally, we also consider the extreme simplification
\eq{
	\label{eq:CIM2}
	r_{\mathrm{CIM2}}(t):= \Delta_c\, \delta_{\pi(t)\, c}\, \epsilon(t)
}
where $\Delta_c$ is a constant. CIM2 stands for Constant Impact Model using two state variables.

Note that the CIM2 is a limiting case of the HDIM2 where $\kappa_{\pi' \pi''}(j) \equiv \Delta_c\, \delta_{\pi'' c}\, \delta_{j\, 0}$. It can also be seen as a limiting case of the TIM2, but in contrast to the latter it is always consistent with the labels $\pi$. Furthermore, the TIM2 is recovered from the HDIM2 letting $\kappa_{\pi' \pi''} \equiv g_{\pi'}$, dropping the sum over $\pi''$. TIM1 is obtained from TIM2 for $g_{\pi'} = g$. Therefore, TIM1 could alternatively be called HDIM1 since the TIM and HDIM ``families'' have the same one-kernel limit. Models with more than two kernels are discussed e.g. in \cite{eisler2011models}.

Despite this formal similarity, transient and history-dependent impact models were previously interpreted quite differently. The conventional view of the TIMs is that $G_\pi(\ell) = \sum_{\ell'=0}^{\ell} g_\pi(\ell')$ is a \emph{transient} response to a single trade that decays over time. It exactly counterbalances the order-sign correlations described in the introduction, such that returns become uncorrelated. The kernels $\kappa_{\pi \pi'}$ in HDIMs, on the other hand, were seen as a means to a \emph{constant} price impact of each trade that depends on the past order-flow. 

In any case, all of the above models are, technically speaking, linear processes adapted to a filtration of the order flow. Therefore, without loss of generality, calling all model returns ``predictions'' seems appropriate in the context of this paper.%
\footnote{
The inclusion of ex post facto labels might appear a bit strange at first, but a clear interpretation is given in the discussion.
}

Each propagator model can be calibrated by solving a linear system of equations expressing the differential price response as the product of a correlation matrix and the desired kernel. The details are explained in appendix~\ref{sec:calibration}. Previously, however, HDIMs were only approximately calibrated by factorising three-point cross-correlations in terms of  two-point cross-correlations. In this work, we introduce a new method which is an extension of the convolution theorem: the full three-point cross-correlations are obtained by performing an inverse two-dimensional Fourier transform on the cross-bispectrum. A self-consistent description is found in appendix~\ref{sec:xxx_proof}. We also calibrated the TIMs using highly efficient spectral estimates for the correlations and response functions. We used a variation of Welch's method by averaging over the cross-correlations calculated independently for each day as described in appendix~\ref{sec:welchpad}. Splitting the time series is actually necessary for the HDIM2 for performance reasons, and also favourable for TIMs to improve the signal-to-noise ratio. In addition, it allows us to perform an out-of-sample analysis by using models calibrated on odd days to predict the returns on even days and vice versa. Therefore, we can rule out deceptive results due to overfitting, which was not guaranteed in previous works. Furthermore, the TIM2 cannot be calibrated without bias using standard methods, as we show in appendix \ref{sec:responses}. We therefore adapt our analyses below to faithfully reveal these biases for the first time.

The CIM2 only has a single parameter $\Delta_c$, which can even be omitted for all analyses of relative scales. Nevertheless, a useful estimate is 
\eq{
\Delta_c = \big\langle |r(t)| \big| \pi(t) = c \big\rangle,
}
which is very close to a half-tick except for very small-tick instruments. The purpose of CIM2 is to test how well price movements can be explained from the instantaneous state $(\epsilon_t, \pi(t))$ alone. 

\section{Calibration results}
\label{sec:results}


\subsection{Time series inspection}

\begin{figure}[h]
  \centering
   \includegraphics{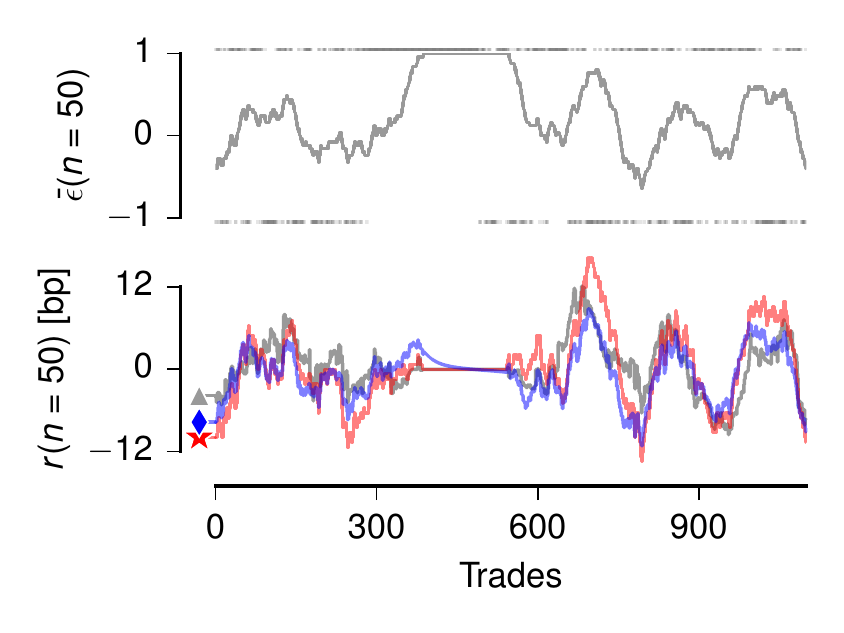}
  \caption{Top: A short excerpt of the order flow for AAPL in 2013, for which $\eta=0.7$. Solid grey line: mean order signs $\bar \epsilon$ in bins of $N=50$ trades (causal). Markers: positions of the trades for each sign. Bottom: grey line starting with a triangle: the true 50-trade return in basis points. Blue line starting with a rhombus:  returns predicted by the TIM2 based the real order flow. The model was calibrated to the whole year. Red line starting with a star: the prediction of the CIM2. Markers are spread out for visibility and connect to the time series, which start at $t=0$. Cross-correlation with the true returns: $0.77$ for TIM2 and $0.74$ for CIM2 (relative estimation errors $\approx 0.05\%$, quarterly variation $\approx 2\%$).}
  \label{fig:pinned_series_AAPL}


 \vspace{\baselineskip}
  \includegraphics{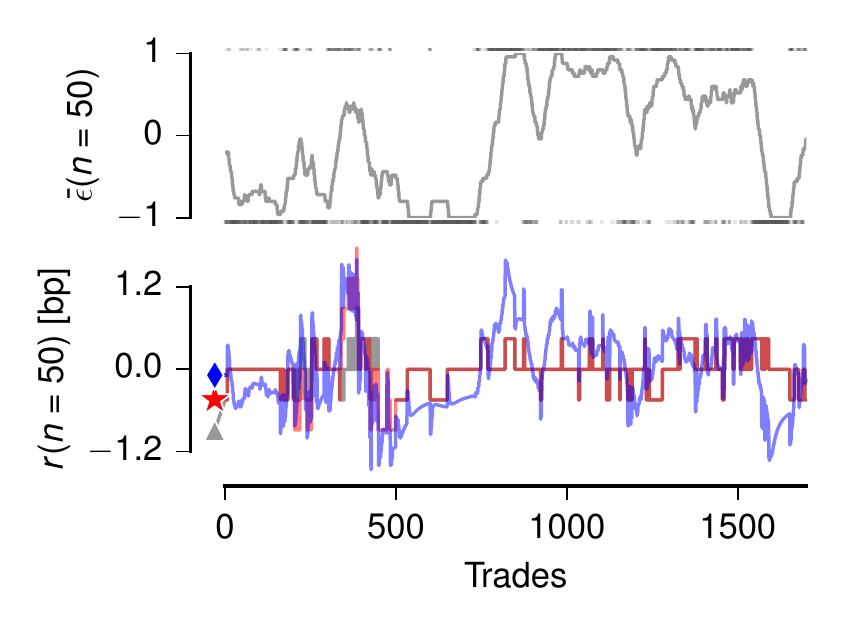}
  \caption{A short excerpt of the order flow (top) and returns (bottom) for SCHATZ in 2013, for which $\eta=0.03$. Methods identical to Fig.~\ref{fig:pinned_series_AAPL}. Cross-correlation with the true returns: $0.51$ for TIM2 and $0.94$ for CIM2 (relative estimation errors $\approx 0.05\%$, quarterly variation $\approx 1\%$). }
  \label{fig:pinned_series_SCHATZ}
\end{figure}

In \cite{patzelt2017universal}, the probability that an order changes the price was identified as a primordial component of aggregate impact. In particular, episodes of extreme order-sign imbalance with prices pinned to a particular level were observed frequently. Fig.~\ref{fig:pinned_series_AAPL} shows a short example of a real order flow for a small tick stock, the corresponding 50-trade return (grey line), and the output of two very different models. In this small-tick example ($\eta=0.7$), the TIM2 (blue line) performs quite well except for a pronounced overshoot when the order-flow becomes very biased. As discussed above, TIM2 is inconsistent with the definition of the label $\pi = n$, since some price movements are predicted in this case. This problem is by construction absent for CIM2 (red line). During episodes of balanced order-flows, the CIM2 performs very similar to the TIM2. Returns from TIM1 are almost identical to TIM2 and HDIM2 combines the best of all models (not shown).

\begin{figure}
  \centering
  \includegraphics{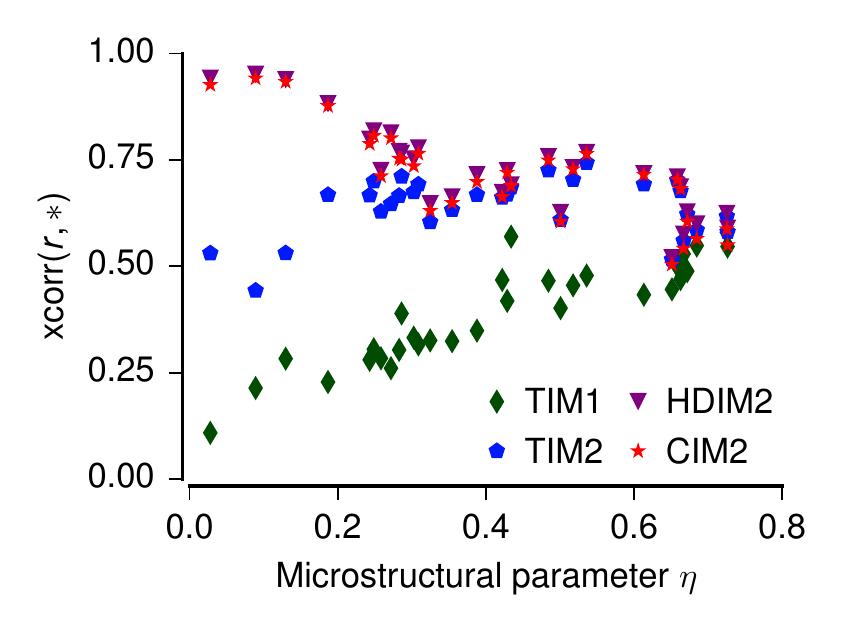}
  \caption{The cross-correlation between the empirical $N$-trade returns ($N=50$) and model predictions based on the true order flow for all instruments in the sample plotted against the corresponding microstructural parameter (see Eq.~\ref{eq:eta}). For each model one point corresponds to one instrument. Model-fits and cross-correlations where calculated for all available 1-year periods in the sample and then averaged if more than one year was available.}
  \label{fig:model_correlations}
\end{figure}

An example for the dynamics of a very large-tick instrument ($\eta = 0.03$) is shown in Fig.~\ref{fig:pinned_series_SCHATZ}. Here, price changes are rare. In consequence, TIM2 tracks poorly and CIM2 very well.

\subsection{Systematic performance comparison}

To compare the models' short-term performances more systematically, the correlation coefficient between real and predicted aggregated returns for $N=50$ successive trades is shown in Fig~\ref{fig:model_correlations}. All instruments in our data set are shown, sorted by their microstructural parameter $\eta$ (see section~\ref{sec:data}). HDIM2 leads to the best result for all values of $\eta$. On average, HDIM2 represents a $2\%$ improvement over CIM2, which is $13\%$ better than TIM2, which is itself $82\%$ better than TIM1. Note, however, that for large-tick instruments CIM2 and HDIM2 models perform almost perfectly while the TIMs are clearly underperforming. For the smallest tick sizes, all models meet at roughly 60\% correlation. The qualitative results are robust with respect to changes of the bin-size $N$, but cross-correlations slowly decline with increasing $N$ for all models.

\begin{figure*}
  \centering
  \includegraphics[width=\textwidth]{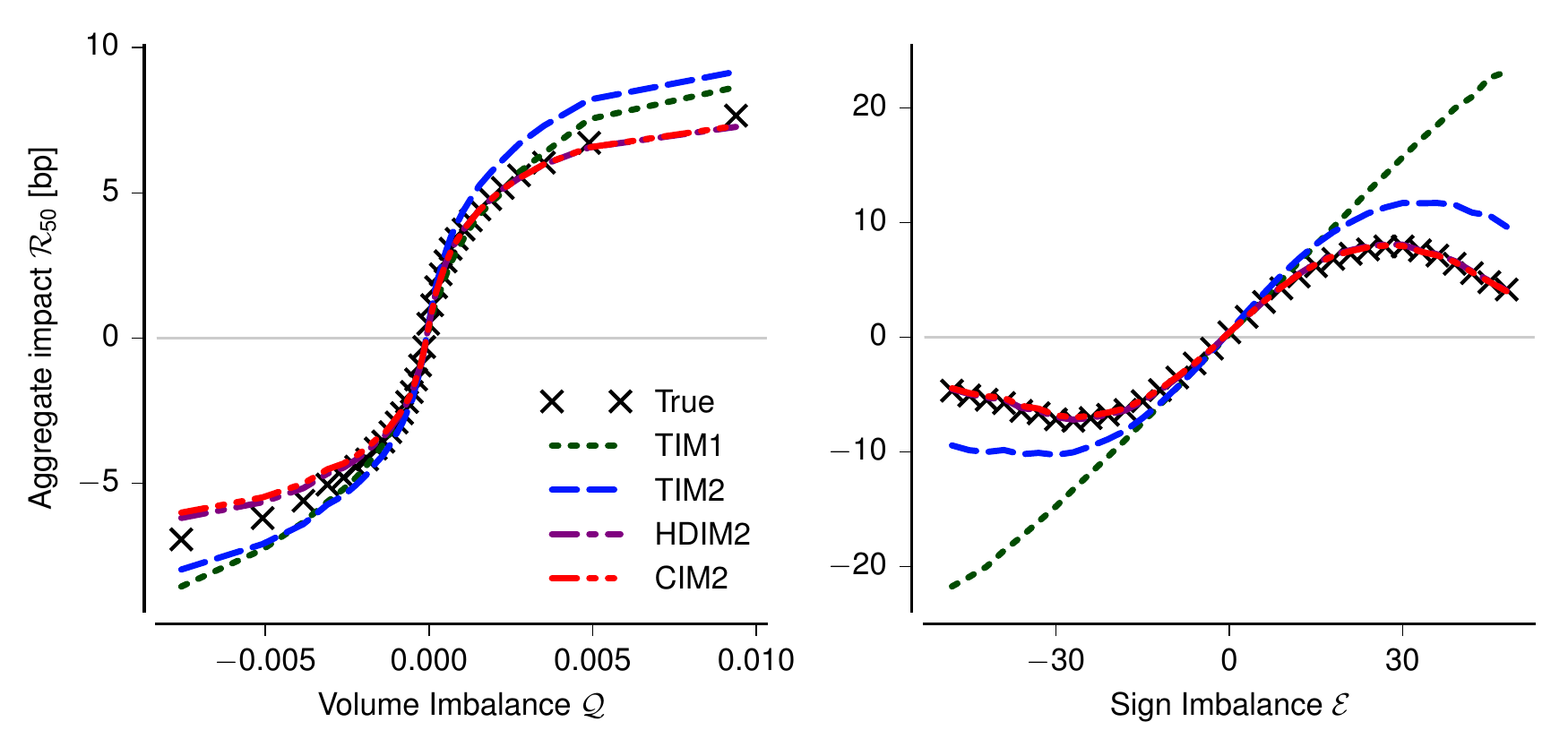}
  \caption{Expected return $\R_N$ conditioned on the aggregate-volume imbalance $\Q$ (left) and on the aggregate-sign imbalance $\E$ (right) in bins of $N=50$ trades. $\Q$ is measured as a fraction of the total daily volume and quartile binned. Impact in basis points. Crosses: measurement for MSFT in 2015-2016. Coloured lines: aggregate impacts for returns predicted by different models given the same order-flow. The curves for the HDIM2 and the CIM2 are almost identical.}
  \label{fig:impact}
\end{figure*}

\begin{figure*}
  \centering
  \includegraphics[width=\textwidth]{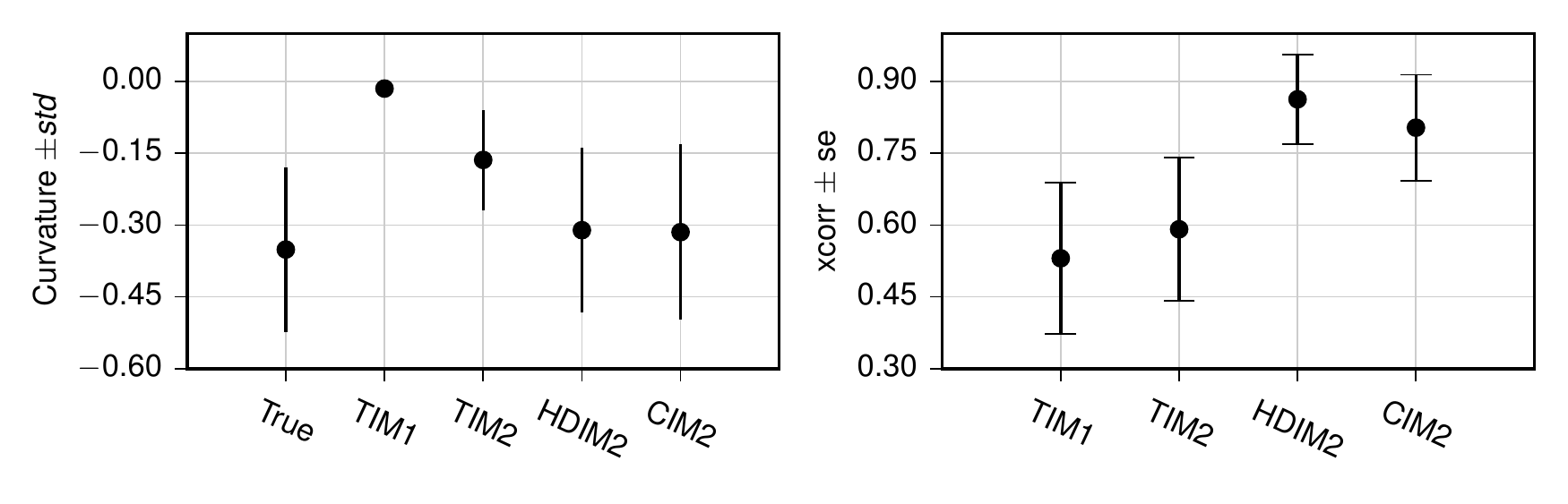}
  \caption{Statistics of the curvature $\chi$ (Eq.~\ref{eq:curvature}) of the sign-impact curves $\R_N(\E)$ for $N=50$, for all instruments. Left: mean curvatures (dots) and dispersion (lines). Right: cross-correlations of the curvatures between model and true impact curves. Error bars: standard errors. }
  \label{fig:curvatures}
\end{figure*}

\subsection{Explaining impact scaling functions}

We now focus on the main quantity considered in \cite{patzelt2017universal}, namely the conditional aggregate impact
\eq{
\R_N(X) := \left \langle \log m({t+N}) - \log m(t)\, \Big{\vert}\, X = \sum_{i=0}^{N-1} x_{t+i} \right \rangle,
\label{eq:aggregate_impact}
}
where $x=q$, the signed volume of single trade, or $x=\epsilon$, the sign of the order. The corresponding aggregate quantity is $X=\Q$ or $X=\E$, respectively.\footnote{
$\Q$ is normalised by total daily volume, which only has a minor quantitative effect.
}

Fig.~\ref{fig:impact} shows $\R_N(\Q)$ and $\R_N(\E)$ for $N=50$ and for a medium tick size stock ($\eta = 0.61$). The sigmoidal shape of $\R_N(\Q)$ is reasonably well reproduced by all models, but HDIM2 and CIM2 are superior to the TIMs. The sinusoidal shape of  $\R_N(\E)$ is, on the other hand, completely missed by TIM1, which leads to an almost linear function. TIM2 fares better, while the HDIM2 and CIM2 reproduce the empirical curve almost perfectly. These results depend on the considered instrument; see also appendix \ref{sec:all_impacts}. For small ticks, all models except TIM1 reproduce the data well. For very large-tick instruments, even TIM2 yields a linear sign-impact function. 

In order to compare quantitatively the models' abilities to reproduce the shape of the aggregate-sign impact curves, we quantify the curvature measure of a curve $f(x)$ on the interval $x \in [-a, a]$ as:
\eq{
	\label{eq:curvature}
	\chi(f) := \frac{1}{3}  - \frac{1}{2} \left( \frac{ \int_{-a/2}^{0}  f(x) dx }{ \int_{-a}^{-a/2} f(x) dx } + \frac{ \int_{0}^{a/2}  f(x) dx }{ \int_{a/2}^{a} f(x) dx }\right).
}
The intuition behind this indicator is that the area under a straight line passing through the origin between $0$ and $a/2$ is one third of the area between $a/2$ and $a$, in which case $\chi = 0$. For a perfect sine function, or for a tent shaped function, $\chi=-2/3$. Being a relative measure based on integration, it performs quite robustly even with noisy data.%
\footnote{We used linear interpolation as an intermediate processing step to be able to integrate over the four intervals despite the irregularly spaced data points.}

Results for $\chi$ averaged over all instruments are shown in the left-hand panel of Fig.~\ref{fig:curvatures}. As reported from \cite{patzelt2017universal}, empirically sign-impact curves $\R_N(\E)$ have a strong negative curvature, $\langle \chi \rangle \approx -0.35$, while $\chi$ is essentially zero for TIM1 and only slightly negative for TIM2. HDIM2 and CIM2, on the other hand, reproduce not only the average curvature but also the standard deviation of the distribution of curvatures across instruments.

The right-hand panel of Fig.~\ref{fig:curvatures} shows the correlations between the predicted and true curvatures of $\R_N(\E)$, across all instruments. As expected, HDIM2 and CIM2  yield the highest cross-correlations, close to 90 \% for HDIM2.\footnote{Note that the square of the cross-correlation can also be interpreted as the coefficient of determination $R^2$ of a linear regression with intercept.} 

\subsection{Diffusivity and Signature Plots}

Another finding in \cite{patzelt2017universal} is that the x-axis and y-axis scalings of the aggregate impact curves are by and large consistent with the Hurst-exponents $H$ of, respectively, the order-sign time series and the return time series. Since we use the former as an input of the model, the only possible discrepancy between models and data can stem from the scaling properties of the output returns.

Previous works studied a very precise measure of deviation from diffusivity, the ``signature plot''
\eq{
	D(\ell) = \frac{1}{\ell} \langle (\log m(t+\ell) - \log m(t))^2 \rangle,
}
which is independent of $\ell$ for perfect diffusion ($H = 0.5$), decaying with $\ell$ for subdiffusion ($H < 0.5$) and growing with $\ell$ for superdiffusion ($H > 0.5$). Here, we follow the literature by subtracting a low-frequency diffusion constant $D_{LF}$ since we are interested only in the deviation relative to perfectly diffusive behaviour.
We obtained $D(\ell)$ by simulating the calibrated models as dynamical systems with the out-of-sample real order-flows as inputs (previous work used closed-form expressions for $D(\ell)$ in terms of in-sample kernels and correlation matrices).

\begin{figure}
  \centering
  \includegraphics{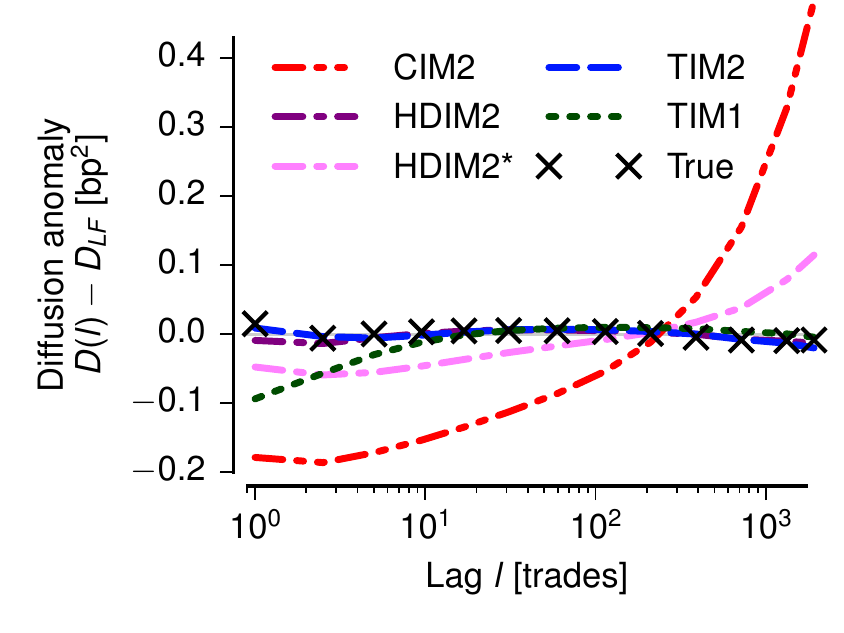}
  \includegraphics{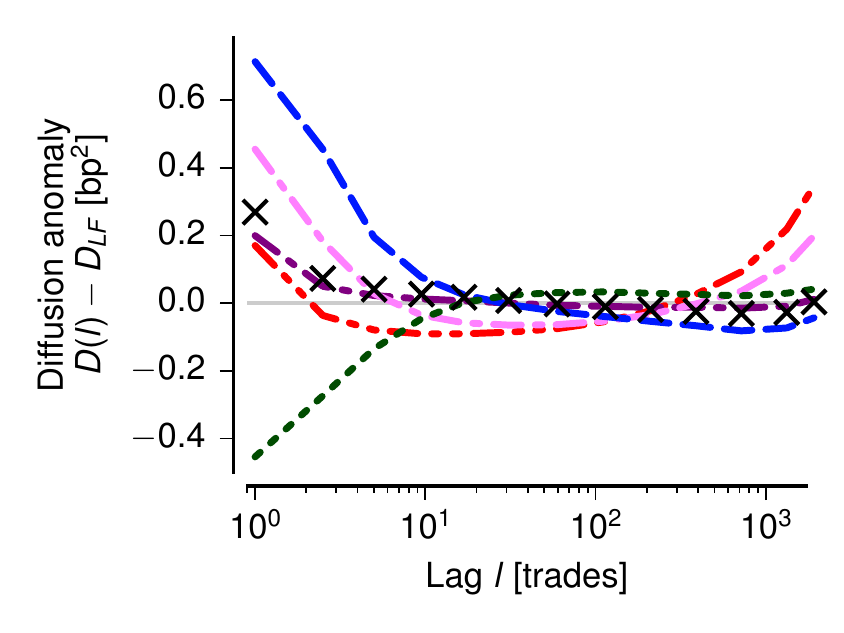}
  \caption{Signature plot for AAPL (top) and MSFT (bottom) in 2015-2016. Crosses: true returns. Coloured lines: aggregate impacts for returns predicted by different models given the same order-flow. For HDIM2* the same two-point approximation as in e.g. \cite{eisler2012price} was used.}
  \label{fig:signature_plot}
\end{figure}

Some examples are shown in Fig.~\ref{fig:signature_plot}. All propagator models tend to behave similarly for small-tick instruments and show more differences for large ticks. For short lags, TIM1 generally exhibits some superdiffusion while TIM2 is strongly subdiffusive for large ticks. CIM2 shows, as expected, superdiffusion for long lags, which is a direct consequence of the long-memory of order sign. As explained in the introduction, this is the reason why propagator models were introduced in the first place.

HDIM2 best reproduces the detailed shape of $D(\ell)$, both for small ticks and for large ticks. This is interesting since previous work, using an approximate calibration of HDIMs based on a two-point factorisation of three-point correlations, reported disappointing results for HDIM2 in the case of small ticks \cite{eisler2011models, eisler2012price, taranto2016linear}. The same effect is seen in Fig.~\ref{fig:signature_plot}, where we show the result of an approximate calibration of HDIM2, that we call HDIM2*. We now see that an exact calibration of HDIM2 resolves this issue, and is able to reproduce accurately both the signature plot and the full response functions, much better than TIM2 (see appendix \ref{sec:responses}).

\begin{figure*}
  \centering
  \includegraphics[width=\textwidth]{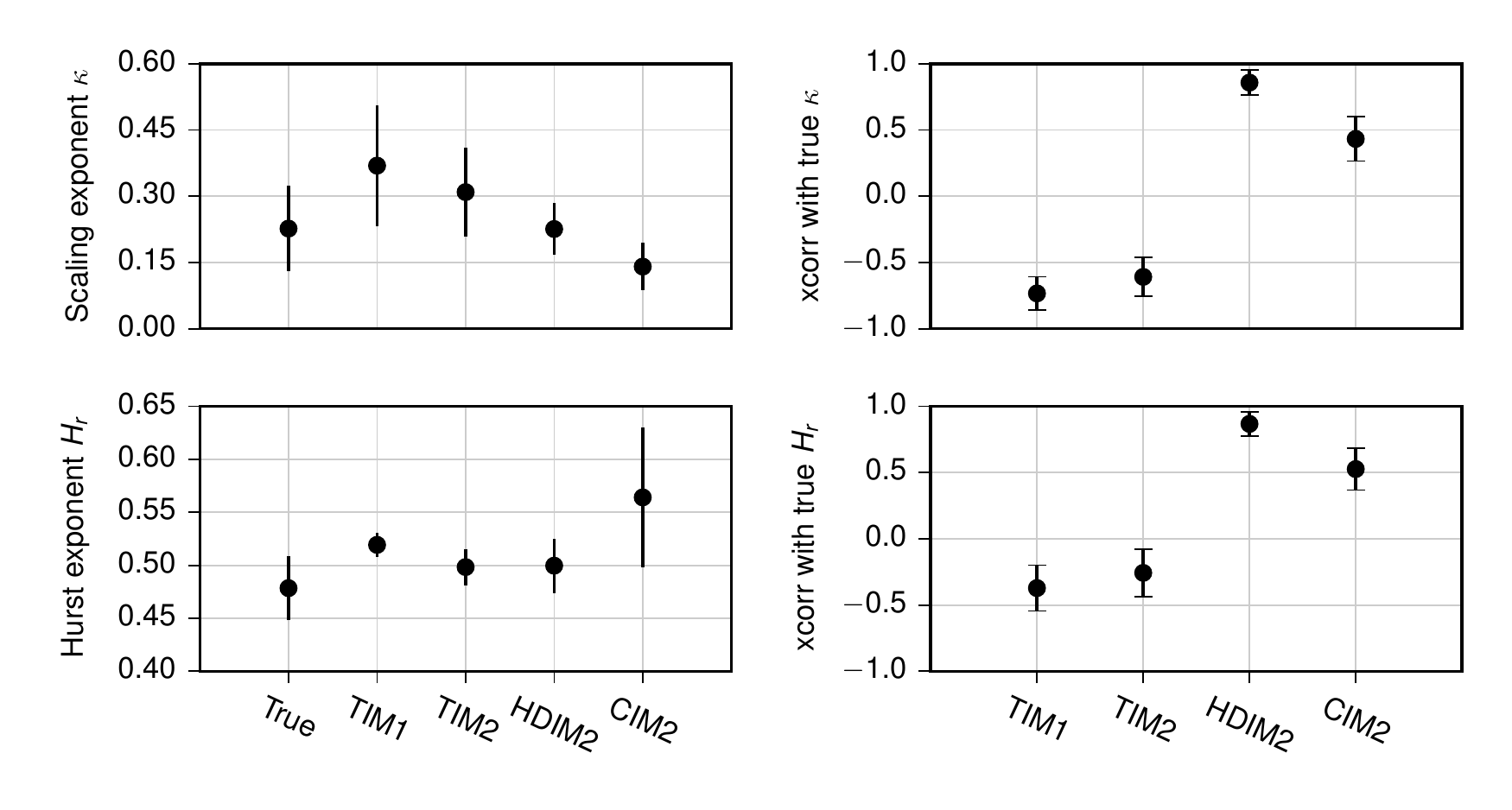}
  \caption{Top left: scaling exponent $\kappa$ for the dependence of the impact curve slope $\partial R_N(\Q)/\partial\Q|_{\Q= 0} \propto N^{-\kappa}$ on the bin size $N$. Markers: means over all instruments for either the true returns (leftmost marker) or the ones predicted by the four models from the true order-flow. Vertical lines: respective standard deviations. Top right: the cross-correlation across instruments of the real $\kappa$ with the model's respective predictions (markers). Error bars indicate standard errors. Bottom left: The Hurst exponents $H_r$ for the same actual and predicted returns used in the upper panel. Means and standard deviation were again calculated across instruments. Bottom right: the cross-correlations across instruments of the real $H_r$ with the models' respective predictions. Error bars: standard errors..}
  \label{fig:scatter_kappa_scaling}
\end{figure*}

A more systematic comparison of the scaling exponents is shown in Fig.~\ref{fig:scatter_kappa_scaling}. The top-left panel shows the distribution across instruments of the scaling exponent $\kappa$ of the \emph{slope} of the impact curves close to the origin, which behaves as $N^{-\kappa}$. HDIM2 again performs best. TIMs tend to overestimate $\kappa$, while CIM2 underestimates it.\footnote{The differences appear to be less pronounced for the aggregate sign impact (not shown)} See appendix \ref{sec:all_impacts} for the full rescaled impact curves for the HDIM2.

The corresponding global Hurst exponents of the return time series' are shown in the bottom-left panel of the same figure. The true returns are slightly subdiffusive (as measured by $H$), while TIM2 and HDIM2  are almost diffusive. The CIM2 is clearly superdiffusive, as expected from the signature plot and the long memory of trade signs, particularly for small-tick instruments. Note, however, that a unique Hurst exponent is unable to describe the full behaviour of the signature plot, see Fig.~\ref{fig:signature_plot}. 

The right hand column of Fig.~\ref{fig:scatter_kappa_scaling} shows the cross-correlations of the slope scaling exponent and the Hurst exponents of the model returns with the real ones. The variations across instruments are almost perfectly reproduced by the HDIM2, for both measures. The CIM2 fares a bit worse in this respect but the TIMs fail completely. Even though their average values (left-hand column) are similar to the true ones, the relative variation across instruments of their long-term diffusivity often moves contrary to the same measures for the true returns.

\section{Discussion}

The main objective of this work was to test whether basal impact models suffice to reproduce the strongly nonlinear, concave shape of aggregate impact functions recently reported in \cite{patzelt2017universal}. In particular, non-price-changing market orders were hypothesised to play a crucial role, which we tested by comparing models that incorporate them in different ways, or not at all. Empirically, price-changes become less likely for more biased order-flows. In consequence, aggregate impact vanishes when the order flow is strongly polarised in one direction \cite{patzelt2017universal}. This apparently counter-intuitive result seems to reflect market conditions where liquidity-takers and -providers selectively act more biased because they cancel out each others impacts.

We revisited several ``propagator'' models proposed in the literature and found that the History Dependent Impact Model HDIM2 is able to accurately reproduce all considered aspects of the dynamics of prices, from the aggregate impact functions to the diffusion properties of the prices. The HDIM2 expresses the next price change as a linear combination of the preceding order signs with weights depending on the respective order-type label--price changing vs. non price changing.

Other models performed worse, particularly the Transient Impact Models. The TIM2 receives the same input as the HDIM2, but it incorporates the different impacts of the two order types only on average, and not consistently for each event. The TIM1 does not distinguish between order types and only considers past order-signs. 

These results led us to investigate how well the various observations can be reproduced just based on the order sign and  the consistent distinction between price-changing and non-price-changing orders at each point in time, without incorporating the preceding order-flow. To this end, we introduced the Constant Impact Model CIM2, where a price-changing event has a constant nonzero impact in the direction of the order-sign while a non price-changing event has zero impact. Although this model has drawbacks--like long-term superdiffusive prices due to the long memory of market order signs--it reproduces the concavity of the aggregate impact functions almost as well as the HDIM2. Furthermore, the assumption that non-price-changing orders have little impact is validated by the weak amplitude of the corresponding kernels (see Fig.~\ref{fig:kernels}).

This shows that non-price-changing orders, which can be caused by taking less liquidity than available and by stimulated refill, are indeed the main cause of the vanishing aggregate-sign impact for biased order flows--simply because long sequences of predominantly buy (or sell) market orders are also long sequences of non-price changing events. It also shows that the nonlinear impact curves are reproduced the HDIM2, which is generally considered a linear model \cite{taranto2016linear}, because the crucial nonlinearity lies in the distinction of whether an order will change the price at all. In this sense, it is a linear model with pre-classified inputs.

Our results were obtained thanks to more rigorous techniques for model calibration and out-of-sample testing than in previous works. For example, we developed a new efficient method to calibrate multi-event HDIMs because HDIM2* calibrated using existing methods from the literature sometimes performs \emph{worse} than its memoryless limit CIM2 (e.g. in Fig.~\ref{fig:model_correlations}, and further data not shown), and than its inconsistent limit TIM2 (see Fig.~\ref{fig:signature_plot}; similar problems were reported before \cite{eisler2012price}). Furthermore, we tested the models by actually running them as dynamical systems and then analysing their output exactly like the true returns. This made our tests more sensitive than the commonly used closed-form expressions. The latter can sometimes conceal model differences because they fail to show calibration biases for inconsistent models like the TIM2. As shown in appendix \ref{sec:responses}, bias-free calibration is not always guaranteed, but it can be verified by measuring the cross-correlation between the model input and its prediction-error. Taken together, our methodological improvements also solved the long-standing question why HDIM models were previously found to often underperform the theoretically less well-founded and sometimes inconsistent TIM models  \cite{eisler2011models, eisler2012price, taranto2016linear}).

Several problems are left for future work. The generation of artificial order-flows was not touched by this paper. This includes the the prediction of non-changing prices, without using this information as an input (see \cite{taranto2016linearII} as a recent attempt). The most important problem, however, might be closing the loop such that order-flow and liquidity dynamics emerge from a recurrent dynamical process.


\FloatBarrier 

\begin{acknowledgments}
We thank M. Benzaquen, G. Bormetti, Z. Eisler,  S. Hardiman,  C.A. Lehalle, I. Mastromatteo, D. Taranto, and B. Toth 
for inspiring discussions. We also thank G. Bolton and J. Lafaye for support during data preparation.
\end{acknowledgments}

\bibliography{impact_models}

\appendix
\section*{Appendices}

\section{Propagator model calibration}
\label{sec:calibration}

Calibrating the TIM and HDIM models to data requires estimating the respective propagator kernels. This amounts to solving a linear system of equations. To see this, consider the (conditioned) differential responses
\al{
	\S_{\pi,\pi'}(\ell)	&:= \langle \delta_{\pi(t)\,\pi'}\,  r(t)\ \delta_{\pi(t-\ell)\,\pi}\, \epsilon(t-\ell) \rangle\\
	\label{eq:s_pi}
	\S_{\pi}(\ell)		&:= \sum\limits_{\pi'} \S_{\pi,\pi'}(\ell)\\
	\S(\ell)			&:= \sum\limits_{\pi} \S_{\pi}(\ell)
}
where $\S(\ell)$ is the expected return one-trade return $r$ at time $t+\ell$ given any trade at time $t$. 

For the HDIM2, combining Eq.~\ref{eq:HDIM2} with Eq.~\ref{eq:s_pi} and introducing the error term
\eq{
	\nu(t) := r(t) - r_\text{HDIM2}(t)
}
yields
\al{
	\S_{\pi}(\ell) &:= \langle r(t)\ \delta_{\pi(t-\ell)\,\pi}\, \epsilon(t-\ell) \rangle\\
	 		   &= \langle (r_\text{hdim2}(t) +  \nu(t))\, \delta_{\pi(t-\ell)\,\pi}\, \epsilon(t-\ell) \rangle\\
	\label{eq:HDIM2_response_approx}
	                   &\approx \langle r_\text{hdim2}(t)\ \delta_{\pi(t-\ell)\,\pi}\, \epsilon(t-\ell) \rangle\\
	\label{eq:HDIM2_response}
			  &=  \sum\limits_{\pi'',\ \pi',\ j \geq 0}\Big[\ \kappa_{\pi' \pi''}(j) \quad \dots \nonumber\\
				&\underbrace{
					\big\langle \delta_{\pi(t) \pi''}\, \delta_{\pi(t-j) \pi'}\, \epsilon(t-j)\, \delta_{\pi(t-\ell) \pi}\, \epsilon(t-\ell) \big\rangle\
				}_{=:\ \C_{\pi \pi' \pi''}(\ell,j) }\Big]
}
with the shorthand notation for the triple cross-correlation $\C_{\pi \pi' \pi''}(\ell,j) := C_{fgh}(\ell,j)$ with $f = \delta_{\pi(t)\, \pi''}$, $g = \delta_{\pi(t-\ell)\, \pi}\, \epsilon(t-\ell)$, and $h = \delta_{\pi(t-j)\, \pi'}\ \epsilon(t-j)$. 

The error $\nu$ vanishes and eq.~\ref{eq:HDIM2_response_approx} holds exactly if and only if the calibration bias
\eq{
	\label{eq:calibration_bias}
	\langle \nu(t)\ \delta_{\pi(t-\ell)\,\pi}\, \epsilon(t-\ell) \rangle = 0.
}
We can expect that a properly calibrated and consistent linear model has no linear correlations between its input and prediction error. Hence $\nu$ should be very close to zero for well-formed models. We discuss when this is true and when problems emerge in section \ref{sec:responses}.

In previous works, $\C_{\pi \pi' \pi''}(\ell,j)$ was approximated by the two-point cross-correlation $\C_{\pi \pi'}(j) := \langle \delta_{\pi(t)\, \pi}\, \epsilon(t)\: \delta_{\pi(t-j)\, \pi'}\, \epsilon(t-j) \rangle$ to simplify calibration. Here we use a new method to calculate the full three point cross-correlation introduced in appendix~\ref{sec:xxx_proof}.%
\footnote{The ordering of the indices is different for $\C$ and $C$ to be consistent with the existing literature. Note that in time series analysis, the cross-correlation is often understood to be normalised by the signals' variances and $C$ would be called a cross covariance. }

Eq.~\ref{eq:HDIM2_response} is a linear system of equations that can be solved for $k$ since $\S$ and $\C$ are observable. In practice, it is useful to rewrite it the form 
\al{
	\label{eq:hdim2_matrix}
	\mS &= \mC\mk \Leftrightarrow\\
	\left[ \begin{matrix} 
		\mS^{nc}\\ 
		\mS^{cc}
	\end{matrix} \right]
	&= 	\left[ \begin{matrix} 
			\mC^{nnc} & \mC^{cnc} \\ 
			\mC^{ncc} & \mC^{ccc} 
		\end{matrix} \right]
		\left[ \begin{matrix} 
			\mk^{nc}\\ 
			\mk^{cc}
		\end{matrix} \right]
}
with the block matrix elements
\al{
	\mS^{\pi\pi'}_{\ell}		&=  \S_{\pi \pi'}(\ell)\\
	\mC^{\pi\pi'\pi''}_{\ell j} 	&=  \C_{\pi \pi' \pi''}(\ell,j)\\
	\mk^{\pi\pi'}_{j}		&=  \kappa_{\pi \pi'}(j).
	\label{eq:hdim2_matrix_elements}
}
We omitted the zero blocks $\mS^{\pi n}$ and $\mk^{\pi n}$. Since $\kappa_{nc}(0)$ is never used and $\S_{nc}(0)$ never observed; their values are arbitrary. The condition $\kappa_{nc}(0) \stackrel{!}{=} 0 $ can be expressed by letting $\mS_0:=0$, $\mC_{0j} := \delta_{j0}$. $k$ can then be obtained using standard numerical solvers.

Removing the respectively unnecessary conditionings, we obtain the analogous system for TIM2
\eq{
		\left[ \begin{matrix} 
			\mS^{nc}\\ 
			\mS^{cc}
		\end{matrix} \right]
	= 	\left[ \begin{matrix} 
			\mC^{nn} & \mC^{cn} \\ 
			\mC^{nc} & \mC^{cc} 
		\end{matrix} \right]
		\left[ \begin{matrix} 
			\mg^{nc}\\ 
			\mg^{cc}
		\end{matrix} \right]
}
with $\mC^{\pi\pi'}_{\ell j} =  \C_{\pi \pi'}(\ell,j)$ and $\mg^{\pi'}_{j} = g_{\pi'}(j)$. 

Finally, the system for TIM1 reduces to $\mS = \mC\mg$ with $\mS_l = S(\ell)$, $\mg_j = g(j)$, and the standard correlation matrix $\mC_{\ell j} = \langle r(t)\ \epsilon(t + \ell - j) \rangle$. Since the latter is a Toeplitz matrix, more optimised numerical solvers are available than for the other models.

In rare cases, we found the estimated kernels too noisy at long lags. This lead to the HDIM2 underperforming the CIM2, e.g. in Fig.~\ref{fig:model_correlations}. We therefore smoothed the estimated kernels without imposing a particular type of decay using logarithmically spaced multiquadratic radial basis functions for lags larger than 10. This improved results significantly such that the HDIM2 never underperforms the CIM2.

\section{Spectral estimation of two- and three-point cross-correlations}
\label{sec:spectral}

\subsection{Proof of the triple cross-correlation estimator}
\label{sec:xxx_proof}

The cross-correlation  $C_{fg}(\ell)$ between \textit{two} functions  $f$ and $g$ at lag $\ell$ is often calculated using the convolution theorem:
\al{
	C_{fg}(\ell) 	&:= \int\limits_{-\infty}^{\infty} \bar f(t) g(t+\ell) dt\\
			&= \F_\nu^{-1}\Big[\bar F(\nu)\, G(\nu)\Big](\ell),
			\label{for:xcorr}
}
where $\F$ denotes the Fourier transform, $\F^{-1}$ the inverse Fourier transform, $F := \F[f]$, $G := \F[g]$, and $\bar f$ denotes the complex conjugate of $f$. 

Analogously, we obtain the triple cross-correlation between \textit{three} time-domain functions $f(t)$, $g(t+\ell)$, and $h(t+j)$ as
\begingroup
\allowdisplaybreaks
\begin{align}
	C_{fgh}(\ell,j)	&:=	\int\limits_{-\infty}^{+\infty} dt\; \bar f(t)\, g(t + \ell)\, h(t + j) \nonumber\\
			  	&=	\int\limits_{-\infty}^{+\infty} dt\;  \BBL \hspace{.31cm}
			 			\int\limits_{-\infty}^{+\infty} d\nu \bar F(\nu) e^{-2 \pi i \nu t}   \quad\dots\nonumber\\
						&\hspace{2cm} \int\limits_{-\infty}^{+\infty} d\nu' G(\nu') e^{2 \pi i \nu' (t + \ell)} \quad\dots\nonumber\\
						&\hspace{2cm} \int\limits_{-\infty}^{+\infty} d\nu'' H(\nu'') e^{2 \pi i \nu'' (t + j)}
					\ \BBR \nonumber\\
				&=	\iiint\limits_{\ -\infty}^{\ +\infty} d\nu\, d\nu'\, d\nu''\, \BBL
					\bar F(\nu)\, G(\nu')\, H(\nu'')\,  \dots \nonumber\\
					&\quad \underbrace{\int\limits_{\ -\infty}^{\ +\infty} dt\, e^{2 \pi i t (\nu' + \nu'' - \nu)}}
						_{=:\,\delta(\nu - (\nu'+\nu''))}
					e^{2 \pi i (\ell \nu' + j \nu'')}  \BBR \nonumber\\
				&=	\iint\limits_{\ -\infty}^{\ +\infty} d\nu'\, d\nu''\, \BBL \bar F(\nu' + \nu '')\, G(\nu')\, H(\nu'')\,  \dots \nonumber\\
					&\hspace{3cm} e^{2 \pi i (\ell \nu' + j \nu'')} \BBR \nonumber\\
				&=	\F^{-1}_{\nu' \nu''} \Big[ B_{fgh}(\nu', \nu'') \Big] (\ell,j),
\end{align}
\endgroup

where $H := \F[h]$ and
\eq{
	\label{eq:bispectrum}
	B_{fgh}(\nu', \nu'') = \bar F(\nu' + \nu'')\, G(\nu')\, H(\nu'')
}
is the cross-bispectrum.

\subsection{Averaging and zero-padding finite discrete-time signals}
\label{sec:welchpad}

In practice, $C(\ell)$, $\ell = -T+1, \dots, T-1$, must be estimated from a finite sample of $T$ observations. For the two-point cross-correlation, this leads to
\al{
	\hat C_{fg}(\ell)	& := \frac{1}{T-|\ell|} \sum\limits_{t=\sup\{0,-l\}}^{T-1-\sup\{0,l\}} \bar f(t)\, g(t + \ell)\\
				& = \text{iFFT}_\nu \big[F(\nu)\, G(\nu)\big](\ell)
}
where $\hat C = \langle C \rangle$ if $f$ and $g$ are jointly wide sense stationary. $F$ and $G$ are the zero-padded and FFTed signals $f$ and $g$. In the following, we always denote cross-correlation estimates by $C$ to simplify notation. 

The normalisation before the above sum depends on the lag $\ell$ because the number of summands decreases with $\ell$. This effect might be negligible for $\ell \ll T$, allowing the use of a biased estimator in many situations. If we average over short segments, however, $\ell$ can become of the same order as $T$. For the same reason, we are padding the signals with $T$ zeros since the output of an unpadded FFT only contains frequencies and phases for $T/2$ positive and negative frequencies, respectively. There are some subtleties with respect to the ordering of the frequencies in the output array and the handling of odd / even $T$, which we don't discuss here since they depend on the particular FFT implementation used.

Now consider the triple cross-correlations. Since both $B_{fgh}(\nu,\nu')$ and $C_{fgh}(\ell, j)$ are $T \times T$ matrices, it is necessary to apply a variant of Welch's method for long signals. That is, to calculate the average over several shorter time-windows. For each segment of length $T$, we find
\eq{
	\label{eq:3xcorr}
	C_{fgh}(\ell,j) = \frac{1}{T-\sup(|\ell|,|j|)}\ \text{iFFT}_{\nu' \nu''} [ B_{fgh}(\nu', \nu'') ] (\ell,j),
}
where $B$ is calculated according to Eq.~\ref{eq:bispectrum} using zero-padded FFT to calculate the Fourier-transformed signals $F$, $G$, and $H$. In the authors implementation, the matrix product in Eq.~\ref{eq:bispectrum} and the inverse 2D FFT in Eq. \ref{eq:3xcorr} are the dominating factors for computational complexity. The iFFT can be performed after averaging all bispectra in some scenarios, such that it becomes negligible for the total performance.

In the notation above, it was attempted to balance precision, readability, and brevity. For the present study, since we calculated the daily correlations for a given instrument before averaging, several complications arose. We chose the number of events on the shortest full trading day (minus one) as the longest lag $L$ considered. For this day, we calculated all transforms after padding with $L$ zeros. The length of each segment to be Fourier transformed was therefore $T=2L$.  For the other days with $L' \leq T$ events, we padded with $T - L'$ zeros. This leads to a straightforward modification of the bias correction discussed above according to the resulting number of nonzero summands.%
\footnote{
Zero-padding also creates a ``blind spot" in each of the off-diagonal quadrants of $C_{fgh}(\ell,j)$ where $|\ell-j| > L$. Fortunately, these values are not needed for the application in this paper.
}
 For each day with more than $T$ events, we calculated the average correlations over the set of segments of length $T$ covering that day with the smallest possible overlap. Finally, we averaged over the results for each day. Therefore, each day has the same weight in the final result, but the contribution of each individual event is lower on very active days. The intuition here is that otherwise, singular days with extreme activities might dominate the result.

\section{Kernels and Responses}
\label{sec:responses}

\begin{figure*}[p]
  \centering
  \includegraphics[width=\textwidth]{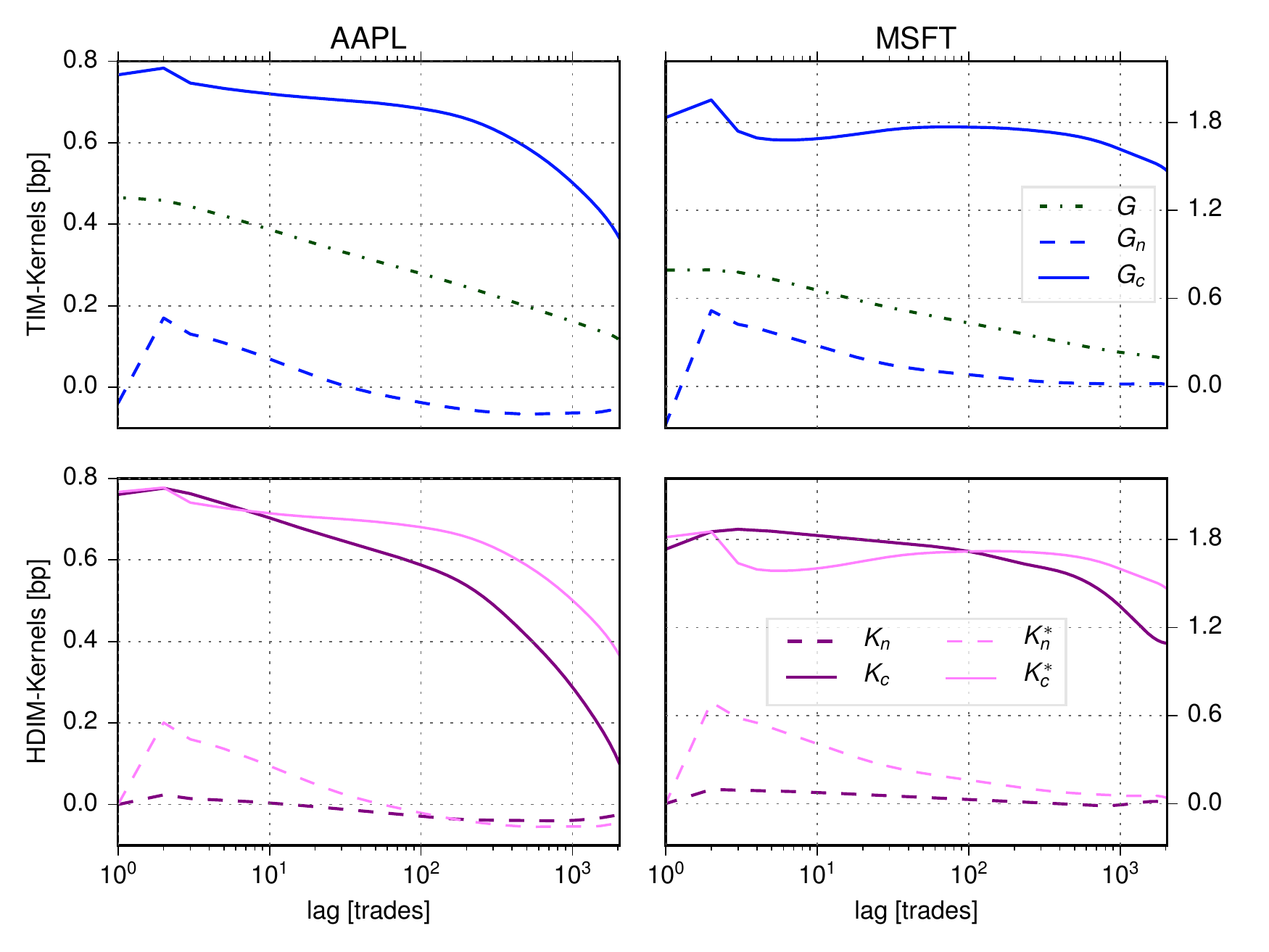}
  \caption{Integrated propagator kernels calibrated to AAPL (left column) and MSFT (right column) for 2015-2016. Top row: ``Bare response'' $G$ for the TIM1, and the conditioned ones $G_\pi(\ell)$ for the TIM2 for $\pi=n$ (non-price-changing events) and $\pi=c$ (price-changing events), respectively. Bottom row: integrated kernels for the HDIM2 calibrated using the two-point approximation $K_\pi^*(\ell)$ and the new complete three-point calibration  $K_\pi(\ell) \equiv K_{\pi c}(\ell) := \sum_{\ell'=0}^{\ell} \kappa_{\pi c} (l')$. }
  \label{fig:kernels}
\end{figure*}

Fig.~\ref{fig:kernels} shows the integrated kernels for the different propagator models. See eq.~\ref{eq:TIM1} ff. and the following explanations, as well as appendices \ref{sec:calibration} and \ref{sec:spectral} for more details. Similar comparisons were done in the past \cite{eisler2011models, eisler2012price, taranto2016linear}, but but this is the first time that an HDIM model was calibrated without approximations. The nonzero kernels for the HDIM2 for both calibrations are shown in the lower row of  Fig.~\ref{fig:kernels}. The approximate kernels (pink lines) are more similar to the TIM2 (blue lines, upper row) than to the full calibration (purple lines). Particularly the impact of non price-changing events (dashed lines) is strongly overestimated for the approximate calibration.

\begin{figure*}[p]
  \centering
  \includegraphics[width=\textwidth]{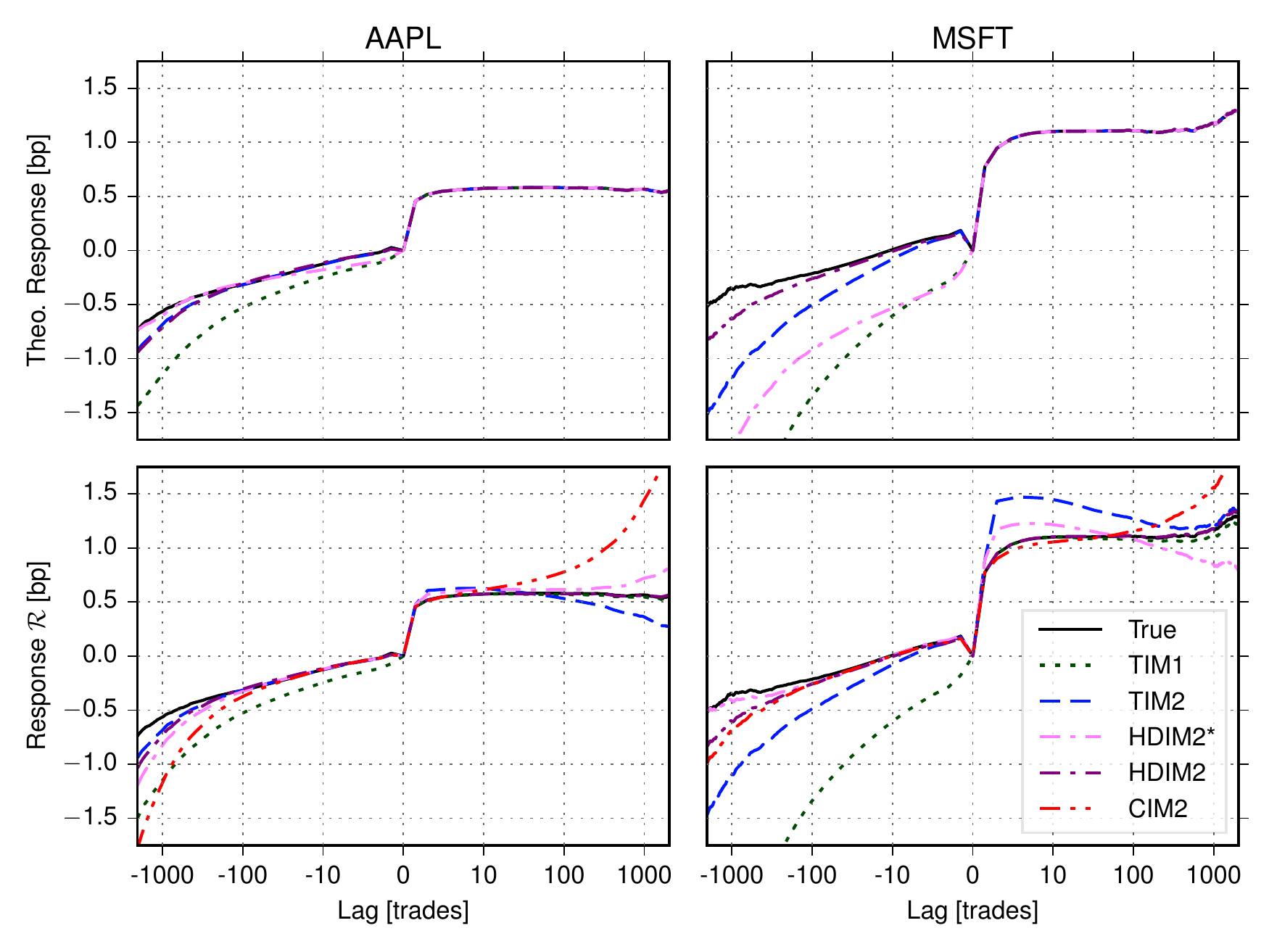}
  \caption{Measured (true) price responses $\R$ and model predictions for AAPL (left column) and MSFT (right column) for 2015-2016. Upper row: model responses calculated from the respective system of equations $\mS = \mC\mk$ or $\mS = \mC\mg$ according to appendix \ref{sec:calibration}. This method is commonly used in the literature. Lower row: ``measured'' model responses by running each model as a dynamical system with the real order-flow as an input. Only this method reveals calibration biases for positive lags.}
  \label{fig:responses}
\end{figure*}

The consistency of an impact model can be checked by considering is the price response 
\eq{
	\R(\ell) := \langle (m(t+\ell) - m(t))\, \epsilon(t) \rangle
}
or the conditioned responses 
\eq{
	\def\arraystretch{1.66}
	\R_\pi(\ell) = \left \{ \begin{array}{lcl}
	 	\sum_{\ell'=0}^{\ell-1} \S_\pi(\ell') \quad & \text{if} & \quad  l > 0\\
		\sum_{\ell'=0}^{\ell+1} \S_\pi(\ell') \quad & \text{if} & \quad  l < 0\\
		0							    & \text{if} & \quad  l = 0
	\end{array} \right.
}
where $\S(\ell)$,  $\S_\pi(\ell)$ are defined according to eqns.~\ref{eq:s_pi} ff. and $\R(\ell) = \sum_{\pi} \R_{\pi}(\ell)$.
Previously, closed form expressions based on the kernels and correlations were the preferred method to calculate price responses (and signature plots) for propagator models (e.g. \cite{eisler2011models, eisler2012price, taranto2016linear}). However, because the true empirical response for positive lags and the correlation functions were used to obtain the kernels, the model responses calculated like this are trivially identical to the measured ones for positive lags. In \cite{taranto2016linear}, the consistency for negative lags was investigated. In the latest revision available as of this writing, however, an approximation was used that actually only involved one element of the kernel for this consistency check.

The predicted responses using the theoretical closed-form predictions based on the full kernels are shown in the upper row in  Fig.~\ref{fig:responses}. By definition, empirical- and model responses are identical for positive lags. The results are therefore potentially deceptive and we only include them because this method (and its the even more flawed variation discussed above) was used predominantly in the literature. For AAPL (left column), all models reproduce also the negative-lag response quite well. Only TIM1 exhibits a noticeably larger discrepancy with too steeply rising (falling) prices before buy (sell) trades. For the larger-tick MSFT (right column), this problem becomes much more pronounced, as discussed in \cite{taranto2016linear}, and affects all models to some degree. The HDIM2 provides the best overall fit, while the approximately calibrated HDIM2* performs much worse.

Actually simulating the models using the real order-flow as an input and then calculating the response functions from the predicted returns produces different results. As we explained before, the calibration bias in eq.~\ref{eq:HDIM2_response_approx} only disappears if there are no correlations between the input and the prediction error of the model. We can expect this to be the case for the TIM1 and HDIM2 since a proper calibration will remove all errors that can be expressed as a linear combination of past inputs. As shown in the lower row of Fig.~\ref{fig:responses}, responses from simulated TIM1- and HDIM2 predictions are indeed almost identical to the the previous method. 

The TIM2, however, deviates significantly from the true responses for positive lags. This is consistent with the overshoot and subsequent reversion observed in Figs.~\ref{fig:pinned_series_AAPL} and \ref{fig:signature_plot}. The cause for this finite calibration bias is the inherent inconsistency of the TIM2: Its output disrespects the event type labels in the input and therefore cannot remove all correlations. We measured the bias according the left-hand side of eq.~\ref{eq:calibration_bias} and found that, as expected, it is close to zero for the HDIM2 but \emph{not} for the TIM2 (not shown). In fact, we found empirically that the difference between the two rows in Fig.~\ref{fig:responses} is exactly explained by this bias (also not shown). Unfortunately, the bias can only be calculated after actually simulating the calibrated model. It therefore allows to verify that a calibration was unbiased \emph{a posteriori}, but not to remove the bias \emph{a priori}.
 
The HDIM2* model also exhibits a number of anomalies. For AAPL, it performs slightly worse using this method and is slightly superdiffusive for positive lags. For MSFT, it exhibits a similar positive-lag overshoot as the TIM2 model. For negative lags, it surprisingly seems to fit the data better, but this is a chance result. It often generates shallower negative-lag responses than predicted using the closed-form method and also compared to other models when they are simulated. Consequently, it occasionally also overshoots the true negative impact significantly for other instruments or time periods (not shown).

The CIM2 shows the typical anomaly for negative lags, and some signs of superdiffusion for positive lags. For large ticks, however, it performs expectedly quite similar to the HDIM2, especially for lags $\ell \in [-100, 100]$.

\begin{figure*}[p]
  \centering
  \includegraphics[width=\textwidth]{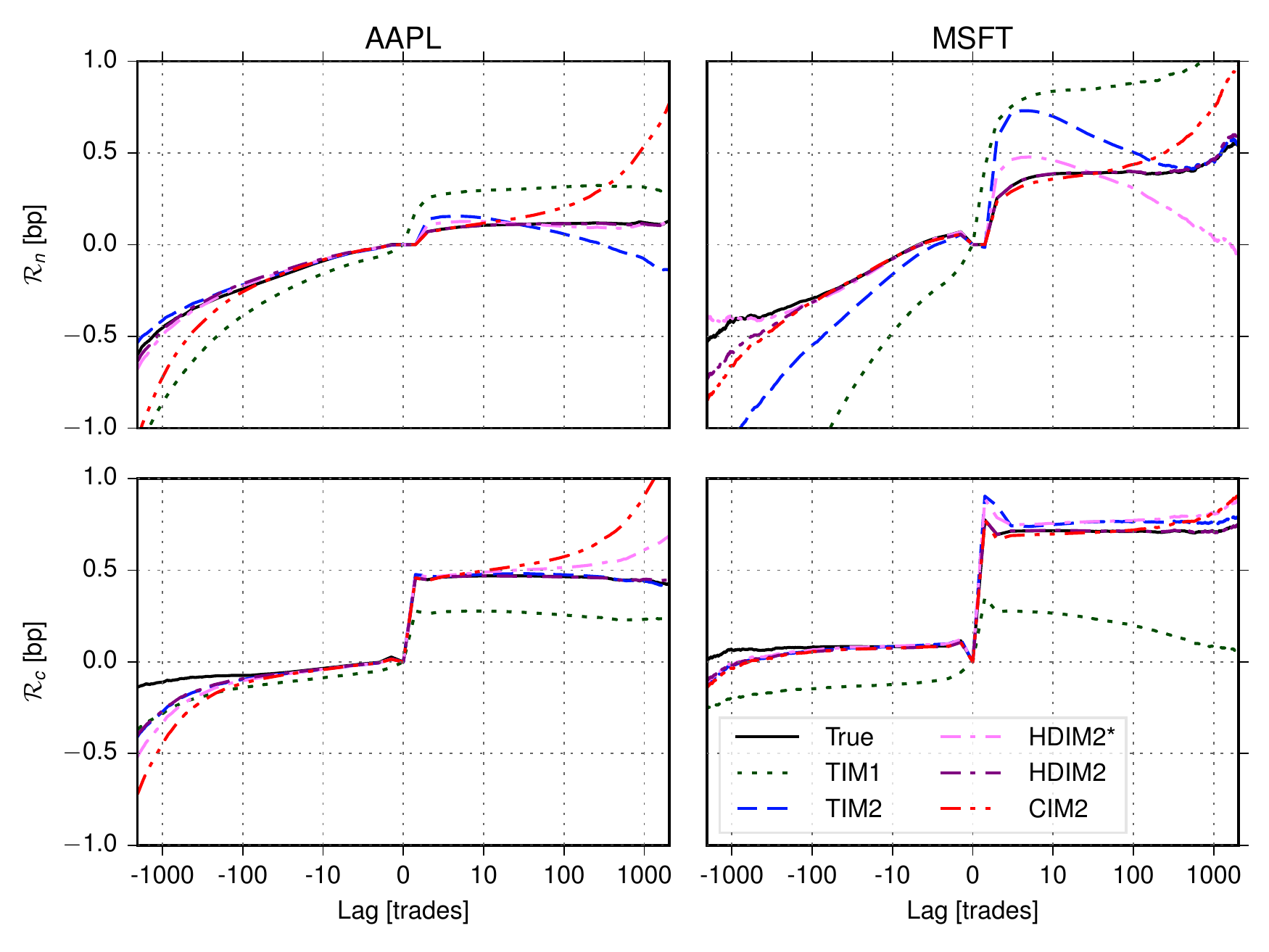}
  \caption{Conditioned price responses $\R_\pi$ for AAPL (left column) and MSFT (right column) for 2015-2016, and for the models simulated using the real order-flow. Upper row: non-price-changing events ($\pi = n$). Lower row: price-changing events ($\pi = c$). }
  \label{fig:responses_nc}
\end{figure*}

Finally, Fig.~\ref{fig:responses_nc} shows the conditioned responses obtained by simulating the respective models. The overshoot of the TIM2 model for short lags is mostly confined to non-price-changing events (upper row). This is consistent with the TIM2 generating price-changes even for non-price-changing events, and compensating for this movement afterwards. For the TIM1, the responses are expectedly far off from the real curve since it doesn't distinguish between the different event labels. The fully calibrated HDIM2 is so close to the true response that it is difficult to see except for $\ell < -100$.

\section{Model impact curves for 30 instruments}
\label{sec:all_impacts}

\begin{figure*}
  \centering
  \includegraphics[width=\textwidth]{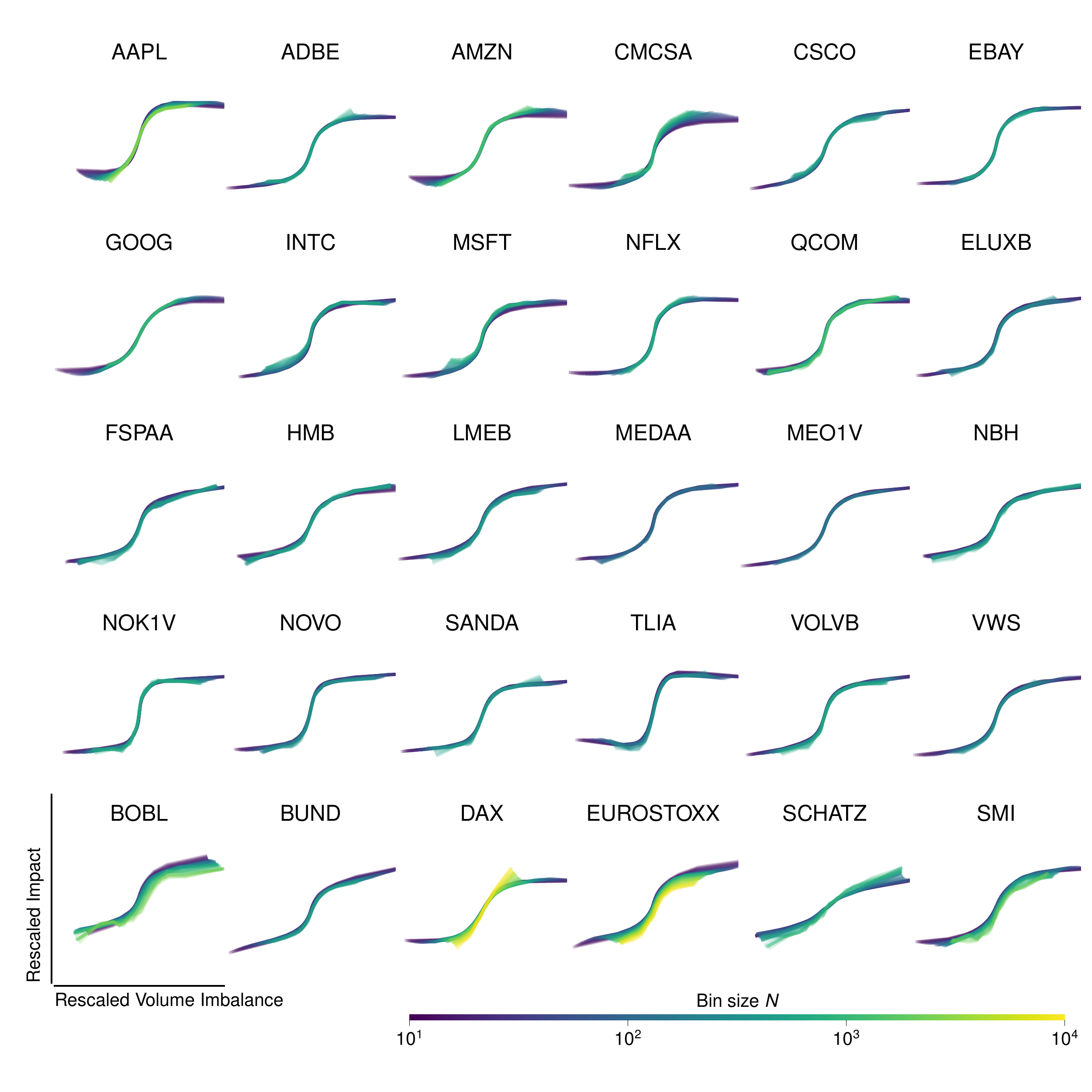}
  \caption{Price impact predicted by the HDIM2 as a function of the true volume imbalance. The raw point-clouds were quantile-binned along the imbalance axis. Therefore the curves have a constant noise level but a range of imbalances that changes with the temporal bin size $N$. First two rows: NASDAQ stocks, 2012-2016. Rows three and four: OMX stocks, 2012-2016. Lowest row: EUREX futures (10/2014-12/2015).}
  \label{fig:rdS-HDIM2-rescaled-volume-impacts-all}
\end{figure*}

\begin{figure*}
  \centering
  \includegraphics[width=\textwidth]{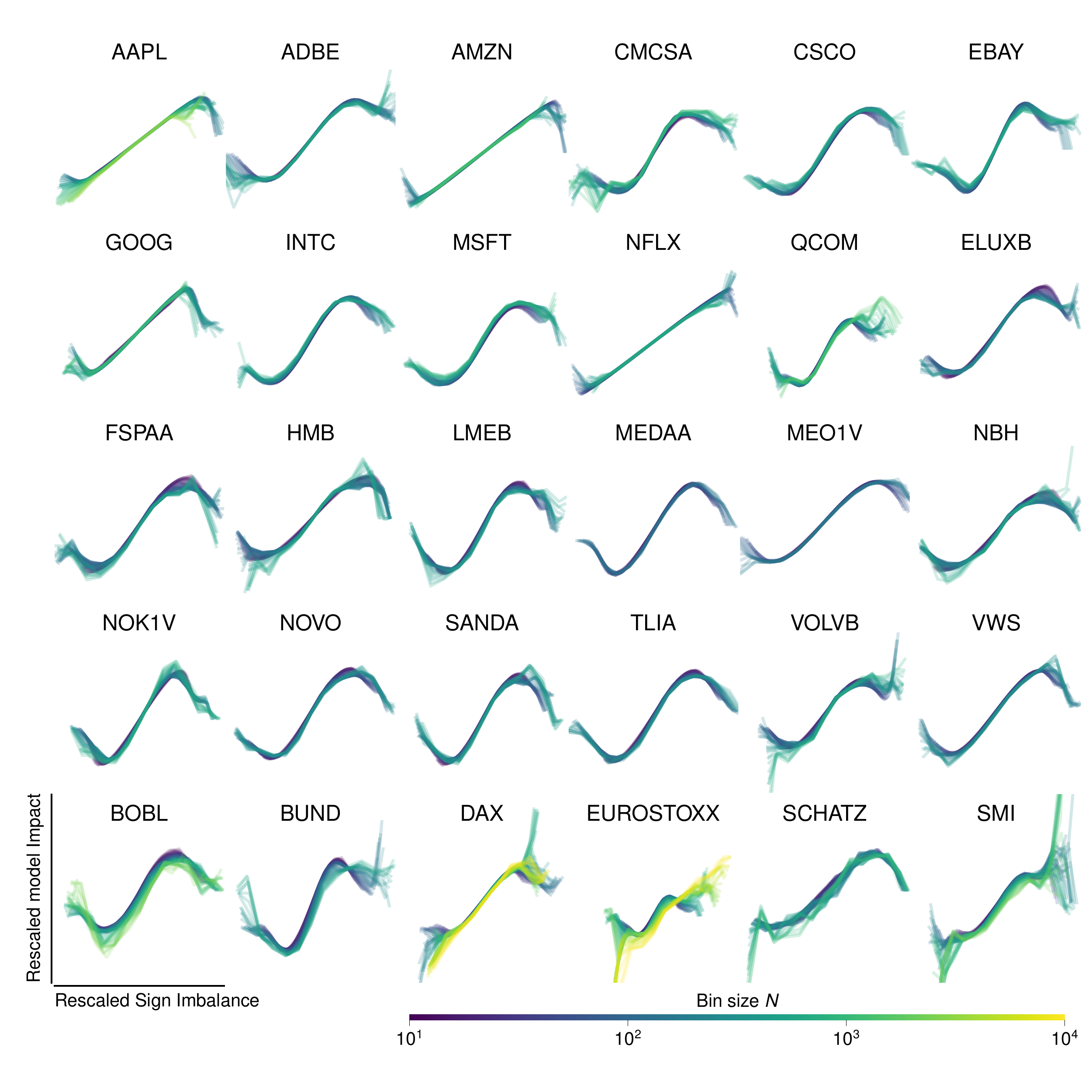}
  \caption{Price impact predicted by the HDIM2 as a function of the true order sign imbalance. Instruments and years as in Fig.~\ref{fig:rdS-HDIM2-rescaled-volume-impacts-all}.}
  \label{fig:rdS-HDIM2-rescaled-sign-impacts-all}
\end{figure*}

Figures \ref{fig:rdS-HDIM2-rescaled-volume-impacts-all} and \ref{fig:rdS-HDIM2-rescaled-sign-impacts-all} show the aggregate-volume and aggregate-sign impact curves, respectively, calculated from the output of the HDIM2 model for all intra-day scales. Visually, they are extremely close to the curves shown in the appendix of \cite{patzelt2017universal} for the true returns. Not shown: the aggregate volume impact curves are quite well reproduced by all models. Also not shown but quantified in Fig.~\ref{fig:curvatures}: aggregate sign curves for the CIM2 model are almost identical to the ones shown here. The ones for the TIM2 are more linear, even perfectly linear for small-tick instruments. For the TIM1, they are always linear.

\end{document}